\documentclass[aps,preprint,superscriptaddress,showpacs,amsmath,amssymb]{revtex4}

\usepackage{graphicx}
\usepackage{bm}
\newcommand{\Op}[1]{{\boldsymbol{\mathrm{\hat{#1}}}}}
\begin{document}
\draft

\title{Quantum Lubrication: Suppression of Friction in a First Principle Four Stroke Heat Engine. }

\author{Tova Feldmann and Ronnie Kosloff}

\affiliation{%
Department of Physical Chemistry,
Hebrew University, Jerusalem 91904, Israel }%

\date{\today}

\begin{abstract}
A quantum model of a heat engine resembling the Otto cycle is employed to 
explore strategies to suppress frictional losses. These losses are caused
by the inability of the engine's working medium to follow adiabatically
the change in the Hamiltonian during the expansion and compression
stages. By adding external noise to the engine frictional losses can be suppressed.
\end{abstract}
\maketitle

\section{Introduction}
\label{sec:introduction}

Working conditions of real heat engines are far from the ideal
reversible limit. Their performance is restricted by 
irreversible losses due to heat transport,  heat
leaks and also friction. Actual working devices tend to optimize the performance by
balancing the losses with maximizing work\cite{salamon01}.
For engines producing finite power irreversible losses are unavoidable. 
High performance engines are therefore constructed from  materials
which reduce heat resistivity while minimizing heat leaks. In
addition lubricants are employed to reduce frictional losses. The
present study explores  {\em quantum lubricants}, schemes to
reduce the {\em frictional} irreversible losses and thus enhance
the performance of the quantum heat engine.

Quantum models of heat engines based on first principles are
remarkably similar to their macroscopic counterparts \cite{kieu04,hoffmann}. 
These engines extract heat from a hot bath of temperature $T_h$ and eject
heat to a cold bath of temperature $T_c$.
The irreversible losses due to the finite rate of heat transport have
been linked to their quantum origin \cite{k85,k87,k116,k122}.
Optimal performance strategies lead to solutions where the working
fluid never reaches thermodynamical equilibrium with the heat
baths. Performance curves can be directly compared to those
obtained in finite time thermodynamics which employ
phenomenological heat transport laws  \cite{curzon75,salamon80}.

Friction is the punishment for compressing or expanding the
working medium too fast. In a quantum engine,
compression/expansion is  a change in an external field described
by a parametrically  time dependent Hamiltonian of the working medium.
Whenever the control Hamiltonian does not commute with the
internal Hamiltonian of the working medium, the rapid change in
the  external field does not allow the state of the working medium to follow
adiabatically the instantaneous energy levels
\cite{k176,k190,k201}. As a result  both coherences and additional energy becomes stored in
the working medium. The dissipation of this additional energy in
the cold bath together with the inevitable decoherence is the
quantum analogue of friction. The key to quantum lubrication is to
suppress  the creation of off diagonal terms in the energy representation.

The quantum four stroke Otto cycle is chosen to demonstrate the lubrication effect.
The working medium is composed from interacting two-level systems. Accordingly, the
uncontrolled internal Hamiltonian becomes:
\begin{equation}
{\Op H}_{int} ~~=~~2^{-3/2} J \left({\Op{\sigma}_x^1} \otimes {\Op{\sigma}_x^2}
-{\Op{\sigma}_y^1}\otimes {\Op{\sigma}_y^2} ~
  \right)~\equiv~J {\Op B_2}~~.
\label{eq:interaction}
\end{equation}
where $\Op \sigma$ represent the spin Pauli operators and $J$ scales the
strength of the inter-particle interaction \cite{k176,k190}. The
external control Hamiltonian is chosen as:
\begin{equation}
{\Op H}_{ext} ~~=~~2^{-3/2} \omega(t)
\left( {\Op \sigma}_z^1 \otimes {\Op I}^2 
+ {\Op I}^1 \otimes {\Op \sigma}_z^2 \right)~\equiv~\omega(t) {\Op B}_1~~,
\label{eq:hext1}
\end{equation}
where $\omega(t)$ represents the external field. The total
Hamiltonian becomes:
\begin{eqnarray}
\begin{array}{c}
{ {\Op H} }
\end{array} ~~=~~
 \begin{array}{c}
\omega(t) {{\Op B}_{1}}+\rm J {{\Op B}_{2}}~~,
\end{array}
\label{matHP}
\end{eqnarray}
where $\Omega (t)= \sqrt{\omega^2+J^2}$ defines the temporary energy scale.
At various times ${\Op H }(t)$ does not commute with itself   since 
$[{\Op B}_{1},{\Op B}_{2}] ~\equiv~ \sqrt{2} i {\Op B}_{3} ~\neq 0~$,
($~{\Op B}_3=2^{-3/2}({\Op{\sigma}_x^1} \otimes {\Op{\sigma}_x^2}
+{\Op{\sigma}_y^1}\otimes {\Op{\sigma}_y^2})$). 
The set of operators $\{{\Op B} \}$ forms a closed orthogonal Lie
algebra. In addition, 
$\left( {\Op B}_k \cdot {\Op B }_j \right) = tr \{ {\Op B}_k^{\dagger} {\Op B}_j \} = \delta_{kj}$ and  $tr \{{\Op B}_k\}=0$. 
The irreversible equations of motion for this set are 
$ ~\frac{d {\Op B} }{dt}~~=~~ i [ {\Op H}, {\Op B} ] + {\cal L_D^*} ({\Op B})$ 
where ${\cal L}_D^*$ is the dissipative Liouville superoperator and the set will be
also closed to ${\cal L}_D^*$. A thermodynamical description
requires that the set of variables 
$\{ {\bf b}(t)\}= \{\langle {\Op B} \rangle \}$ 
should be closed to the dynamics generated on all
branches of the engine's cycle.

The energy balance of the engine is composed of the heat flow and power:
\begin{equation}
\frac{dE}{dt}= {\cal P} + \dot {\cal Q}
\label{eq:dedt}
\end{equation}
where: $ {\cal P} ~~=~~\langle\frac{ \partial \Op H}{\partial t}
\rangle = \dot \omega \langle {\Op B_1} \rangle~~~, $ and $ \dot
{\cal Q} ~~=~~ \langle {\cal L}_D^* ( {\Op H}) \rangle$.

The state of the working medium $\Op \rho$ can be reconstructed from 
five thermodynamical variables $b_{\bf k} =\langle {\Op B}_k \rangle $ composed of the expectation
of the three operators in the Lie algebra and two additional ones, 
$\Op B_4~=~2^{-3/2} \left(  \Op {\sigma}_z^1 \otimes {\Op I^2}
- {\Op I^1}  \otimes \Op {\sigma}_z^2 \right)$
and $\Op B_5~=~\frac{1 }{ 2}{\Op {\sigma}_z^1} \otimes {\Op {\sigma}_z^2}$
\cite{k190} leading to:
\begin{equation}
{\Op \rho}~~=~~\frac{1}{N} {\Op I} + \sum_{\bf k} b_{\bf k} {\Op B_k}~~~. 
\label{eq:dens}
\end{equation}

The occupation probability $p_n$ of the energy level $n$ defines the energy
entropy:
\begin{equation}
{\cal S}_E ~~=~~ -\sum_n p_n \ln p_n ~~.
\end{equation}
If $[{\Op \rho} , {\Op H }]\neq 0$ then this entropy is different from
the von Neumann entropy
\begin{equation}
{\cal S} ~~=~~ - tr \{ \Op \rho \ln \Op \rho \}~~,
\label{eq:entropyv}
\end{equation}
and ${\cal S}_E \geq {\cal S}$. The difference between ${\cal S}_E$
and ${\cal S}$ is a signature of {\em friction} \cite{k201}.
The external entropy production is a measure of the irreversible
dissipation to the hot and cold baths:
\begin{equation}
\Delta { S}_{cyl}^{ext} ~~=~~ -\left( \frac{ {\cal
Q}_h}{T_h}+\frac{ {\cal Q}_c}{T_c} \right) ~~\geq ~0~,
\label{eq:entrprod1}
\end{equation}
where ${\cal Q}_{h}$ and ${\cal Q}_{c }$ are the heat dissipated to the hot or cold baths
respectively.

\section{The Cycle of Operation}
\label{sec:ocycle}

\begin{figure}[bt!]
\includegraphics[scale=0.3]{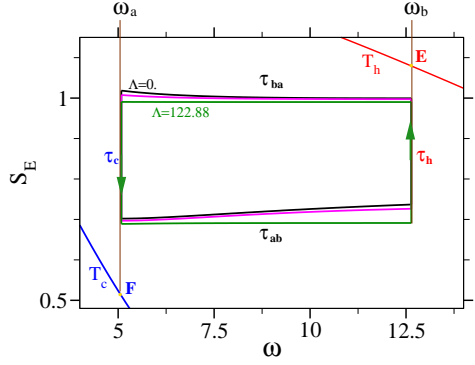}

\caption{Three cycles of the heat engine in the ($\omega, {\cal S}_E$) 
plane. The common values are :$ ~J=2.,~T_h=7.5,~T_c=1.5,
  ~~\omega_a=5.08364,~ \omega_b=12.6355,
 ~\Gamma_h=~ \Gamma_c=1.16748,
~\gamma_h=-0.05, ~\gamma_c=-0.06$.  
$\Lambda_{ba}=\Lambda_{ab}=0$  
for the black cycle,~ $\Lambda_{ba}=~1.28,~\Lambda_{ab}=~ 0.64$ for the
red cycle , and ~~$\Lambda_{ba}=122.88,~\Lambda_{ab}=61.44.$ for the
green cycle. The time allocations on the different branches correspond 
to the optimal power point, when  
$\Lambda_{ba}=\Lambda_{ab}=0$. The  
optimal time allocations are: $\tau_h=1.0795~ \tau_{ba}=0.01478,~\tau_c=1.0088 ~,~
\tau_{ab}=0.0069$. The points $\bf E$ and $\bf F$ represent the equilibrium 
points of the hot and cold {\em isochores}.} 
\label{fig:3cycles}
\end{figure}

A four stroke cycle of operation is studied. As shown in  Fig. \ref{fig:3cycles} 
this cycle includes:
\begin{itemize}
\item{An {\em adiabatic} expansion branch where  an external field
is chosen to decrease linearly from $\omega_b$ to $\omega_a$ } \item{A cold  {\em
isochoric} branch where heat is transferred from the working
medium to the cold bath ($T_c$).} \item{An {\em adiabatic}
compression branch where  an external field is increased linearly from
$\omega_a$ to $\omega_b$.}\item{A hot  {\em isochoric} branch
where heat is transferred from the hot bath at temperature $T_h$
to the working medium.}
\end{itemize}
This cycle is a quantum model of  the macroscopic Otto cycle.
The control parameters are the time allocations on the
different branches, the total cycle time and the extreme values of
the external field.

The  cycle of the engine becomes a sequence of four completely
positive maps that define the different branches. Eventually this
sequence closes upon itself. Repetition of the sequence of
controls leads to steady state operational conditions or a limit
cycle \cite{k201}. The map ${\cal U}_k$ relates the initial set of
these operators to their final values for each of the engine
branches. These maps are obtained by solving the equations of
motion for the set of operators $\{\Op B \}$. The overall cycle
map is the product of the individual maps of each branch 
$ {\cal U}~~=~~  {\cal U}_{ba}{\cal U}_{c} {\cal U}_{ab}{\cal U}_{h}$
\cite{k190,k201}.

On the {\em isochores } the maps ${\cal U}_{h/c}$ are generated by
the completely positive generator  ${\cal L}_D^*$  \cite{lindblad76}. 
Forcing detailed balance ${\cal L}_D^*$ leads to thermal equilibrium with a rate
determined by $\Gamma$.
In addition ${\cal L}_D^*$ also degrades the off diagonal elements of 
$\Op \rho_e$, interpreted either as decoherence or as dephasing. 
The dephasing time $T_2$ becomes identical to the energy equilibration
time $T_2=T_1$. The dissipation has to also eliminate the additional energy
accumulated on the {\em adiabat}. Degrading the coherences
causes the {\em frictional} process to become irreversible \cite{k190,k201}.
The interaction of the working medium with the bath can also be
elastic. These encounters will scramble the phases conjugate to
the energy, and the associated decay time is termed pure dephasing
($T_2^*$). In Lindblad's formulation it becomes ${\cal
L}^{\ast}_{D^e}(\Op B)=-\gamma [\Op H,[\Op H, \Op B]]$ and
$T_2^*=1/2 \gamma \Omega^2$. Note
that elastic medium can not transfer or absorb heat, therefore we
always need an inelastic medium.

On the {\em adiabats}, the varying field $\omega(t)$ causes an explicit time dependence
of $\Op H$ $\frac{ d{\Op B}}{dt} ~~=~~ i [ {\Op H(t)}, {\Op B}]$. 
Since the energy eigenvalues change, even if initially $[{\Op \rho} ,{\Op H}]=0$ 
the state $\Op \rho$ will develop off-diagonal terms in the energy frame (Cf. Eq. (B6) in Ref. \cite{k201}).
The external power ${\cal P}= \dot \omega b_1$ can also be decomposed to the diagonal and off-diagonal
terms in the energy representation:
\begin{equation}
{\cal P} =  \dot {\Omega}\frac{\langle {\Op H} \rangle}{\Omega}  + 
\frac{{\dot \omega} J}{\Omega^2}
\langle \left(J {\Op B_1} - \omega  {\Op B_2}\right) \rangle ~~~.
\label{eq:power4}  
\end{equation}
The first diagonal term represents the power required to {\em compress} or {\em decompress} 
the working fluid:
\begin{equation}
{\cal P}_{ad}^{field}~~=~~\dot {\Omega}\frac{\langle {\Op H} \rangle}{\Omega}~~,
\label{eq:pext}
\end{equation}
the second term in Eq. (\ref{eq:power4}) is the additional power required to drive the working fluid
in a finite rate:
\begin{equation}
{\cal P}_{ad}^{friction}~~=~~\frac{{\dot \omega} J}{\Omega^2}
\langle \left(J {\Op B_1} - \omega  {\Op B_2}\right) \rangle ~~.
\label{eq:pfric}
\end{equation}
This term represents the power invested against friction therefore it vanishes when $J=0$ or $\dot \omega=0$ \cite{k152}.

\section{Quantum Lubrication}
\label{sec:lubrication}

A good lubricant should be able to increase the overall optimal power of the engine.
The insight that energy coherences leads to frictional losses,
suggests that forcing the cycle trajectory to follow adiabatically
the  instantaneous energy levels will be beneficial. A possible
approach is to increase the dephasing on the {\em isochores} so
that at the beginning of the {\em adiabat} $[\Op \rho, \Op H]=0$.
A thermal bath with strong dephasing will cause such an effect.  We have
found that an addition of pure dephasing has only a minor effect
on the performance of the engine. 

The quantum "lubricant" has to suppress the creation of the energy coherences 
on the {\em adiabats}. Formally this can be described by a generator of dephasing  in
the equations of motion for the set $\{ {\Op B} \} $ on the {\em adiabat}:
\begin{equation}
\frac{d {\Op  B}}{dt}~~=~~ i[{\Op H},{\Op B}]~~-~\Lambda[\Op H, [ \bf \hat H, \Op B]]
\label{eq:dephad1}
\end{equation}
and $\Lambda$ is the dephasing coefficient. 

\begin{figure}[bt!]
\includegraphics[scale=0.3]{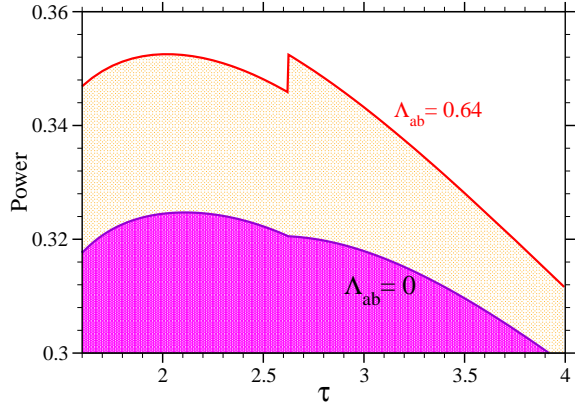}
\caption{The optimal power as function of cycle time $\tau$ with and without
the quantum "lubricant". The optimal time allocations are obtained
for $\Lambda_{ba}= \Lambda_{ab}=0$ \rm(blue line). The red line is the power
with $\Lambda_{ba}=~1.28 \Lambda_{ab}=~ 0.64$.~
All the other parameters are as in Fig. \ref{fig:3cycles}. 
}
\label{fig:dephopt}
\end{figure}

The success of this approach is shown in Fig. \ref{fig:dephopt}. As a reference the optimal
power of the engine as a function of cycle time is shown around the global maximum. Each point on the graph is optimized with respect to the time allocations on the four branches of the cycle. Employing these time allocations the power of the engine is recalculated with the addition
of the dephasing term on the {\em adiabats}. 
It is clear in Fig. \ref{fig:dephopt} that in the interval of cycle times 
around the maximum power the "lubricated" engine outperforms the optimal 
solutions of the reference engine. The "lubricated" maximum power point also moves to 
shorter cycle times. 

For longer time allocations on the {\em adiabats}  where less external power is consumed to overcome the friction,
the performance enhancement due to dephasing decreased eventually leading to a crossover where dephasing 
on the {\em adiabats } decreased the power. For larger  $J$ values we also found that dephasing was not able to 
improve the performance.

\begin{figure}[bt!]
\includegraphics[scale=0.3]{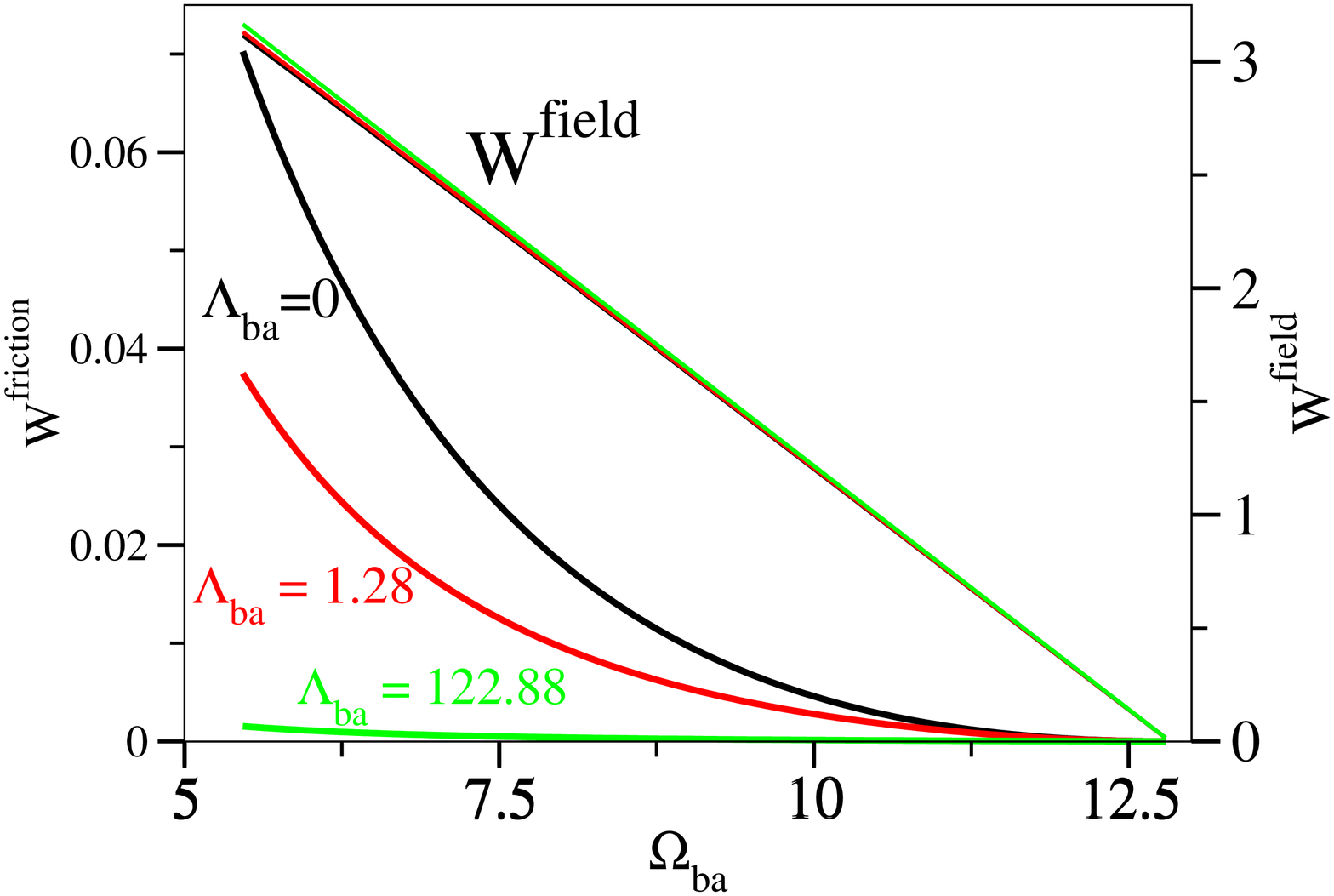}
\caption{The work performed by the engine on the {\em adiabat} $b \rightarrow a$ branch,
separated into $W^{friction} $ (left scale) and $ W^{field}$ (right scale) 
for different dephasing values
as a function of $\Omega_{ba}(t) $.
$ W^{field}$ has only a weak dependence on the dephasing value.
The cycle parameter values are as in Fig. \ref{fig:3cycles}.}    
\label{fig:workfric} 
\end{figure}

Fig. \ref{fig:workfric} shows the accumulated work against friction 
${\cal W}^{friction} = \int  {\cal P}^{friction} dt
= \int  \frac {\dot { \omega}J(b_1J-\omega b_2)}{\Omega^2}dt$ (Cf. Eq. (\ref{eq:pfric})~)
as a function of time on the adiabat for increasing dephasing parameter. 
The main point is that increasing dephasing  eliminates the work against friction. 
This improvement  saturates once ${\cal W}^{friction}$ is eliminated.

Another consequence of the quantum lubrication is that the energy entropy ${\cal S}_E$
does not increase on the {\em adiabats} as can be seen in Fig. \ref{fig:3cycles}.
As a result the energy entropy ${\cal S}_E$ approaches the von Neumann entropy 
${\cal S}$, Eq. (\ref{eq:entropyv}).
These results establish the principle of
quantum lubrication maintaining the working fluid in a diagonal state
in the energy representation.

\subsection{Dephasing synthesis}

Suppression of friction requires a method to synthesize 
dephasing  on the {\em adiabats}. The dynamics of the working medium has to be changed from
unitary to dissipative. The obvious approach of adding a dissipative bath
on the {\em adiabatic} branches is difficult to achieve. Such a bath should have only elastic
encounters with a system with a time dependent Hamiltonian.

The solution is to employ the external controls of the engine 
to synthesize the dissipation.
The idea comes from the singular bath limit, a bath generated from a system operator
coupled to a delta correlated noise ${\Op H}_{sr}= {\Op A_s } g(t)$ 
where $\langle g(t) g(t') \rangle_r = \gamma \delta (t-t')$ where the average 
is taken over the bath fluctuations. The Liouville generator associated with this system bath
coupling becomes: ${\cal L}^*({\Op X})~~=~~ -\frac{\gamma^2}{2} [ {\Op A_s} ,[{\Op  A_s} , {\Op X} ]]$
\cite{gorini76,ingarden75}.

To implement such a scheme random noise is added to the external controls of the engine.
The implementation divides the {\em adiabat} branch into $N$ segments. In each of these segments,
the external field $\omega$ is constant and is chosen to be (for the $a \rightarrow b~$ {\em adiabat}): 
$\omega_k= \omega_a + \frac{ \omega_b-\omega_a }{N} k $ for the $k$th segment.
The short time propagator on the $k$ segment for the set $\Op B$ becomes:
\begin{equation}
{\cal U}_k {\Op B}~~=~~  e^{i {\Op H}(\omega_k) \Delta t_k} ~ 
{\Op B} ~e^{-i {\Op H}(\omega_k) \Delta t_k}~~,
\label{eq:kprop}
\end{equation}
where $\Delta t_k$ is the time interval of the $k$th segment.
At this point random noise is added to the time interval 
\begin{equation}
\Delta t_k ~~=~~ \frac{\tau_{ab}}{N}(1 + r)
\end{equation}
where $r$ is a random number with zero mean and variance $\sigma$. Expanding
the propagator Eq. (\ref{eq:kprop}) to second order and averaging over 
the random noise will lead to the average generator for the $k$ time segment: 
${\cal L}_k^* ({\Op B})= i [  {\Op H}(\omega_k),{\Op B}]- \frac{N \sigma^2}{2 \tau_{ad}} [{\Op H}(\omega_k),[  {\Op H}(\omega_k),{\Op B}]]$. 
In the limit when $N \rightarrow \infty$ this average propagator becomes 
identical to Eq. (\ref{eq:dephad1}) provided $\sigma = \sqrt{\frac{2 \tau_{ad} \Lambda}{N}}$.

The addition of random noise means that the individual cycle has to be
replaced by the average performance on many cycles. As a result
only an average cycle time can be defined. 
This noisy lubrication procedure was simulated with $N=200$
on both {\em adiabats}. The power and other thermodynamic variables were
calculated as an average of 2000 cycles. Convergence was checked by continuing this averaging
1000 additional times. 

\begin{figure}[bt!]
\includegraphics[scale=0.3]{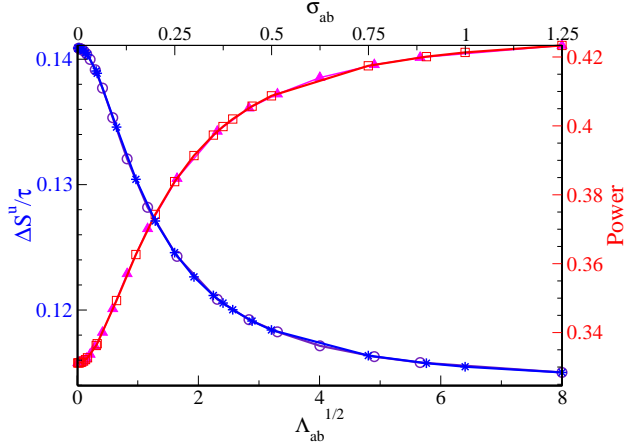}
\caption{Power (right scale) and entropy production (left scale)
as a function of the dephasing parameter $\Lambda_{ab}^{1/2}$  and 
the variance, $\sigma_{ab}$  of the random fluctuations of the
time segment on the {\em adiabats} (upper scale). The dephasing synthesis results are represented by
stars and filled triangles. The empty circles and squares represent the dephasing dynamics.
The calculations were performed with $\frac { \Lambda_{ab}  }{\Lambda_{ba}}=\frac { \sigma_{ab}  }{\sigma_{ba}}$. 
}
\label{fig:dephpods} 
\end{figure}

Fig. \ref{fig:dephpods} compares the Power ${\cal P}$ and entropy production
per cycle, $\Delta S^u/\tau$  calculated by the two methods for the 
time allocations of the maximum power point. It is clear that the results obtained by 
imposing dephasing on the adiabats Eq. (\ref{eq:dephad1}) are identical to the 
dephasing synthesis Eq. (\ref{eq:kprop}).
The signature of lubrication is  the reduction of entropy production which accompanies 
the increase in power. 
This is contrary to optimizing the power with respect to heat transport. In that case the increase 
in output power is accompanied by an increase in entropy production \cite{k152}.
The choice of the procedure to generate dephasing is unique. For example,
adding the random noise to the frequency $\omega_k$ at each time segment has been tested.  
The performance of the engine only became worse.
The reason is that such a term leads to the dissipative generator 
${\cal L}_D({\Op A}) = -\frac{\gamma}{2} [\Op B_1,[\Op B_1, \Op A]]$ which 
does not eliminate the off diagonal elements in the energy representation.

The present study should be related to other recent works.
For example, adding mechanical noise  to a quantum refrigerator has been shown
experimentally to cool  atoms in a magnetic trap  \cite{kumakura03}. 
It seems that the mechanism involves inducing non unitary dynamics. 
Contrary to the present study, in other scenarios
coherence can be beneficial.
Without violating the second law Scully et. al. \cite{scully03}
showed that additional work can be extracted from the coherences
in quantum heat engine.

To summarize, frictional losses are caused whenever 
$[ \Op H_{ext},\Op H_{int}] \ne 0 $. Then the fast dynamics
induces coherences in the energy frame.  
The essence of quantum lubrication is suppressing the generation of 
these off diagonal elements in the energy
representation.  
The present model demonstrates how externally induced noise
can achieve this task.

\begin{acknowledgments}
Work supported by the Israel Science foundation. 
We want to thank Lajos Di\'osi David Tannor and Peter Salamon
for many helpful discussions and Moshe Goldstein
for the random number generator.

\end{acknowledgments}

\end{document}